\documentclass[sigconf]{acmart}
\AtBeginDocument{%
  }

\setcopyright{acmlicensed}
\copyrightyear{2026}
\acmYear{2026}
\setcopyright{cc}
\setcctype{by}
\acmConference[CIKM '26]{Proceedings of the 35th ACM International Conference on Information and Knowledge Management}{November 07--11, 2026}{Rome, Italy}
\acmBooktitle{Proceedings of the 35th ACM International Conference on Information and Knowledge Management (CIKM '26), November 07--11, 2026, Rome, Italy}
\acmDOI{10.1145/3799682.3840177}
\acmISBN{979-8-4007-2539-5/2026/11} 

\usepackage{placeins}
\usepackage{subcaption}

\begin{document}

\title{The EMN Country Factsheets Structured Dataset}

\author{David Alonso del Barrio}
\email{ddbarrio@idiap.ch}
\affiliation{%
  \institution{Idiap Research Institute}
  \country{Switzerland}
}

\author{Daniel Gatica-Perez}
\email{gatica@idiap.ch}
\affiliation{%
  \institution{Idiap Research Institute and EPFL}
  \country{Switzerland}
}
\renewcommand{\shortauthors}{Alonso del Barrio et al.}
\begin{abstract}
\footnote{\textbf{David Alonso del Barrio, Daniel Gatica-Perez| ACM 2026. This is the author's version of the work. It is posted here for your personal use. Not for redistribution. The definitive Version of Record will be published in CIKM'26 ACM International Conference on Information and Knowledge Management https://doi.org/10.1145/3799682.3840177}}
Each year, the European Migration Network (EMN) country factsheets deliver an overview of key migration and international protection developments within all EMN Member States and observer countries. The factsheets include both a textual component and a visual component. In this paper, we introduce a curated dataset of the textual component of these reports over 35 countries and 13 years (2012-2024.) The dataset was created to facilitate European-level research on migration policies and developments, and promote the use of reliable sources about migration in data science and media research, particularly at a time when the spread of online misinformation about migration constitutes a serious issue. The dataset transforms the original document texts into a tabular format, with columns corresponding to country, year, section, subsection, content, and harmonized title section. 
We illustrate the value of the dataset with concrete analyses and propose envisioned applications and uses of the dataset. The dataset is accessible through a DOI link.
\end{abstract}



\maketitle

\section{Introduction}
Migration is a fundamental issue for Europe,
with significant media coverage \cite{eberl2021mapping}. It is a topic that generates polarization and opposing narratives, which can lead to mistrust and doubts in society about what true information is \cite{seiger2025navigating}. In the context of development and policy, this polarization creates a critical need for ground-truth 
and institutional baseline information that reflects official government actions and communication strategies, rather than public or media speculation \cite{boswell2011role}.

In Europe, one of the main actors in the generation of reliable migration data is the European Migration Network (EMN) \cite{mastenbroek2022political}. EMN is a network of National Contact Points (NCPs) composed of national network stakeholders with expertise in migration. EMN NCPs are designated by their respective national governments and are housed in a variety of institutions, including Ministries of Interior and Justice, specialized government agencies focused on migration, research institutes, non-governmental organizations, and national offices of international bodies \cite{EMN2024AMO}. 

One of the main roles of the EMN NCPs is to generate reports on migration in their respective countries \cite{emn2018leaflet}. 
In our work, we built a dataset composed of the textual content of the EMN factsheets. These documents are shorter versions of extensive annual reports. The text in the factsheets is useful for describing and contextualizing policies and developments in the area of migration, and provides high-level context for making comparisons between countries.  

In the migration field, several projects aim to evaluate migration policies through indicators, where experts from various countries first evaluate individual policies through questionnaires, and then provide scores based on different indicators \cite{Solano2022}. For example, the project Immigration Policies in Comparison (IMPIC) \cite{helbling2017measuring, helbling2018migration, berger2024immigration} measures admission policies (border control, entry criteria, etc.) of 33 OECD countries during the period 1980-2018. Another example is the Migrant Integration Policy Index (MIPEX) \cite{yavcan2025mipex}, which assesses integration policies (such as labor market access, education, and citizenship paths) against a standard of equal rights across 56 countries, relying on national experts to evaluate policy indicators from 2007 to the present. A third approach is the one by the Determinants of International Migration (DEMIG) \cite{de2014compiling, de2015conceptualizing} and Quantifying Migration Scenarios for Better Policy (QuantMig) projects \cite{schreier2023demig}, which track specific policy events or changes over time, rather than a static index score. This approach is designed for research into the long-term evolution and effectiveness of policy changes, allowing researchers to study cause-and-effect relationships over decades.

Unlike indexes such as MIPEX or IMPIC, which prioritize numerical scoring, or DEMIG, which tracked policy changes, our dataset provides a standardized reporting framework of official administrative actions. Since these factsheets are authored by NCPs, they represent a unique form of state self-representation: they are not only facts, but the official narrative of how a government justifies and structures its migration management. The dataset thus captures the process by which governments define migration challenges, assign responsibilities, and articulate solutions that they consider normative for their society \cite{banulescu2021we}.

This dataset complements existing research in two ways. First, it provides the qualitative context often missing from purely statistical indicators. Second, it aligns with the field of "Text-as-Data" in political science \cite{grimmer2013text}, where texts such as speeches, social media, or policy documents are treated as raw data to be analyzed using quantitative methods and natural language processing (NLP). While recent studies have used computational text analysis to track the evolving narratives of the International Organization for Migration (IOM) and the  Office of the United Nations High Commissioner for Refugees (UNHCR) \cite{green2025talking}, there remains a data gap for systematically analyzing how European governments report on migration. 

The curated longitudinal and cross-country dataset enables a comparative analysis of how institutional language and policies evolve. By offering a structured version of these documents, our work supports both policy evaluation, by cross-referencing these reports with external indices, and data science applications, such as using official government data as a ground-truth baseline for disinformation and media framing studies.

Additionally, the dataset includes historical texts (2012–2019) that are not currently available in public digital archives, thus preserving a critical period of European migration history.

\section{Data description}
\label{sec:data_description}
EMN factsheets are divided into two main parts, one textual and one visual. In the textual part, each country explains the main points related to topics like asylum, legal migration, integration, or irregular migration, while in the visual part there is a statistical annex that complements the textual part with plots and statistics. Our dataset is focused on the textual part. This part is divided into sections. 

The sections are defined by titles like \textit{"LEGAL MIGRATION"}
\textit{"BORDERS, SCHENGEN AND VISA"} or
\textit{"TRAFFICKING IN HUMAN BEINGS"}, and all these titles are part of column \textit{"section"} in our dataset.  Over the years, the titles of these sections in the textual part have slightly changed. For example, the titles presented above appear with some variations in certain years:\textit{ "LEGAL MIGRATION AND MOBILITY"}, \textit{"BORDERS AND VISAS"}, or \textit{"ACTIONS AGAINST TRAFFICKING IN HUMAN BEINGS"}. These variations in the titles motivated the creation of an additional column called \textit{"grouped\_title\_section"}, enabling consistent comparisons of related topics regardless of year-to-year small title variations. Inside each section, there are cases with subsections  (e.g., section: \textit{"LEGAL MIGRATION AND MOBILITY"}, subsection: \textit{"FAMILY REUNIFICATION"}). All these subsections are part of column \textit{"subsection"} in our dataset. In order to keep the structure of the dataset, in the cases of sections without a subsection, we used the term \textit{"Main"} as the name of the \textit{"subsection"}. Then, the column \textit{"content"} contains the specific content for that section and subsection, and the columns \textit{"year"} and \textit{"country"} specify the corresponding year and country of the factsheet. 

\begin{figure}[htp]
  \centering
\includegraphics[width=0.6\linewidth]{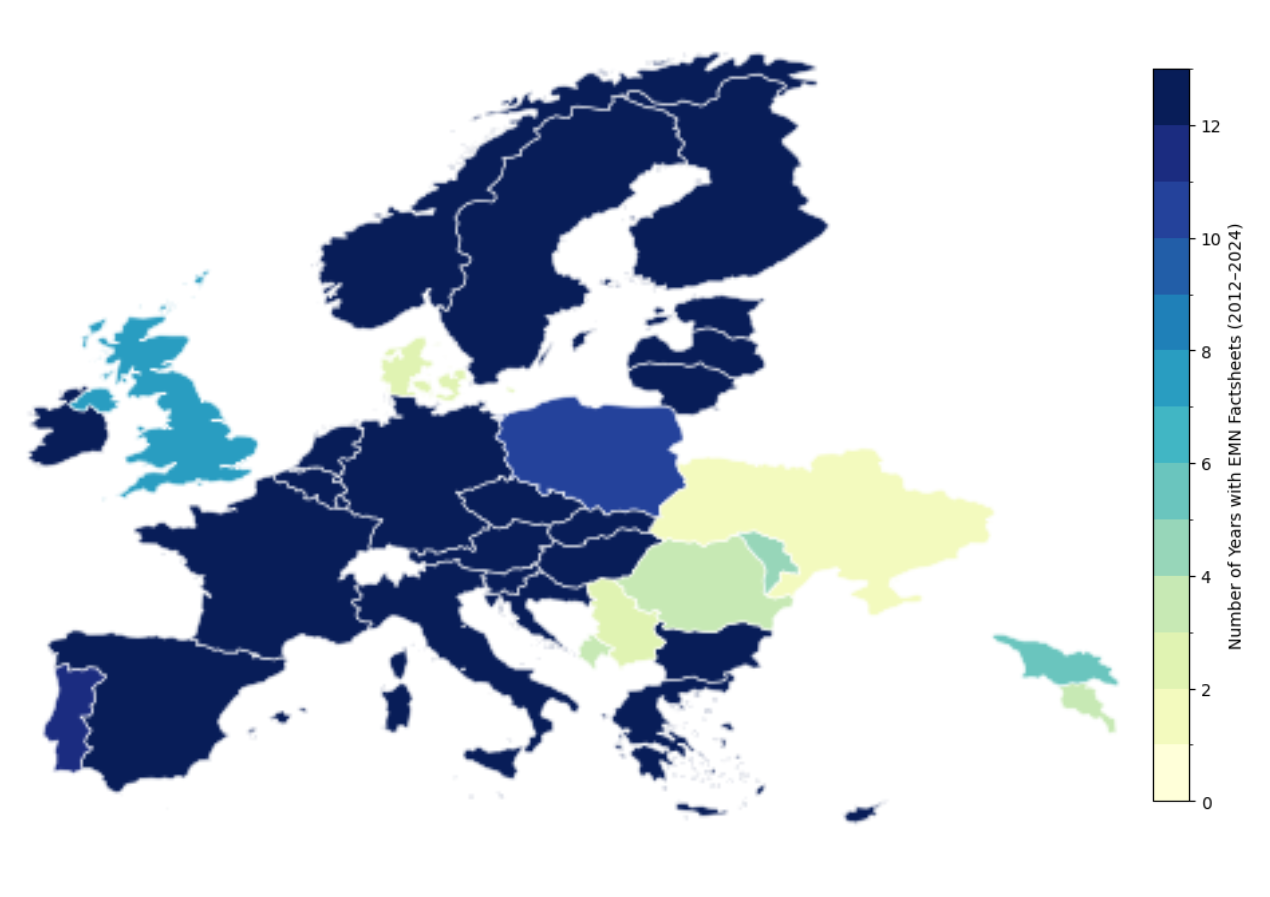}
  \caption{Geographic coverage of the EMN factsheets textual dataset across Europe for the period 2012--2024.
        }
\label{fig:map}
\end{figure}

Figure \ref{fig:map} shows the countries and the number of years in which they have published EMN factsheets. It is important to mention that not all countries have data in all years. For example, the United Kingdom reported to EMN for the period 2012-2018, but it stopped reporting afterwards due to its departure from the EU. 
There are other countries like Serbia (2023, 2024), Montenegro and Armenia (2022, 2023, 2024), or Ukraine (2024) that began to report in recent years, or the cases of Denmark (2012, 2013) and Romania (2012, 2013 and 2014), which only reported in specific years.
 
 
 Table \ref{tab:summary} shows an example of both one sample and the total number of countries (35), years (13), sections (37), subsections (99), content (4801) and grouped\_title\_section (15) in the dataset. 

\begin{table}[!htp]\centering
\resizebox{7.5cm}{!}{%
\begin{tabular}{|c|c|c|}
\hline
\textbf{Columns} & \textbf{Example} & \textbf{Total} \\
\hline
country & Spain & 35 \\
\hline
year & 2020 & 13 \\
\hline
section & BORDERS, SCHENGEN AND VISAS & 37 \\
\hline
subsection & VISA POLICY & 99 \\
\hline
content & \parbox{7cm}{
The visa sections of Spanish embassies and consulates were not closed at any time due to COVID-19. C visas were issued in cases exempt from the entry ban, in accordance with Orders passed by the Ministry of the Interior. D visas were issued normally, as entry with a national visa was allowed at all times.
} & 4801 \\
\hline
grouped\_title\_section & BORDERS AND VISAS & 15 \\
\hline
\end{tabular}%
}
\label{tab:summary}
\caption{Structure and example content of the dataset.}
\label{tab:summary}
\end{table}

\section{Data Collection and Curation Process}
\label{sec:data_collection}
The EMN website\footnote{\url{https://home-affairs.ec.europa.eu/networks/european-migration-network-emn/emn-publications/country-factsheets_en}} provides access to EMN factsheets for recent years. We downloaded all available documents from 2020 to 2024. In addition, we contacted the European Commission (EC), who provided us archival factsheets from 2012 to 2019, resulting in a corpus of 359 PDF documents. 

The collection presents significant heterogeneity in document structure and encoding. We developed an automated pipeline to handle two primary formats. 
\textit{Standard Extraction}, for PDFs with embedded text, where we utilized the PyMuPDF library \footnote{\url{https://github.com/pymupdf/PyMuPDF}}, and 
\textit{OCR-based Extraction}, for documents where text was rendered as images (observed primarily in the 2017–2020 period). For the last case, we implemented an OCR layer using pytesseract \footnote{\url{https://pypi.org/project/pytesseract/}}.

In some cases, factsheets exhibited hybrid structures, containing both machine-readable text and image-based sections, requiring specific interventions to ensure data integrity. Additionally, the automated extraction introduced artifacts, such as OCR typos, malformed characters, and inclusion of document metadata (e.g., footers, page headers, and superscript footnote markers) that were not part of the core content. To address this, we performed a rigorous multi-pass manual curation process, reviewing the dataset three times to correct errors and standardize formatting.

We further harmonized the extracted text, restructuring it into a schema of 6 columns (country, year, section, subsection, content, and grouped\_title\_section). This resulted in the final version of the structured dataset.

To ensure transparency and reproducibility, our extraction and harmonization pipeline is publicly available in a GitLab repository\footnote{\url{https://github.com/idiap/emn-factsheet-pipeline}}. This repository provides representative samples from 2018 and 2022 to illustrate our workflow.

\section{Illustrating the Value of the Dataset}
\label{sec:example}
 We now illustrate the structural richness, longitudinal depth, and thematic diversity of the data. By examining cross-country variation, section-level heterogeneity, and selected thematic trajectories, we highlight how the dataset enables comparative policy analyses. 
\subsection{Dataset Descriptive Analysis}
With the aim of better understanding how each country contributes, whether there are significant differences between countries, or how different the various sections that make up the factsheets are in terms of length, we show two plots that provide a summarized view of the content.
\begin{figure}[htbp]
    \centering
    \includegraphics[width=\linewidth]{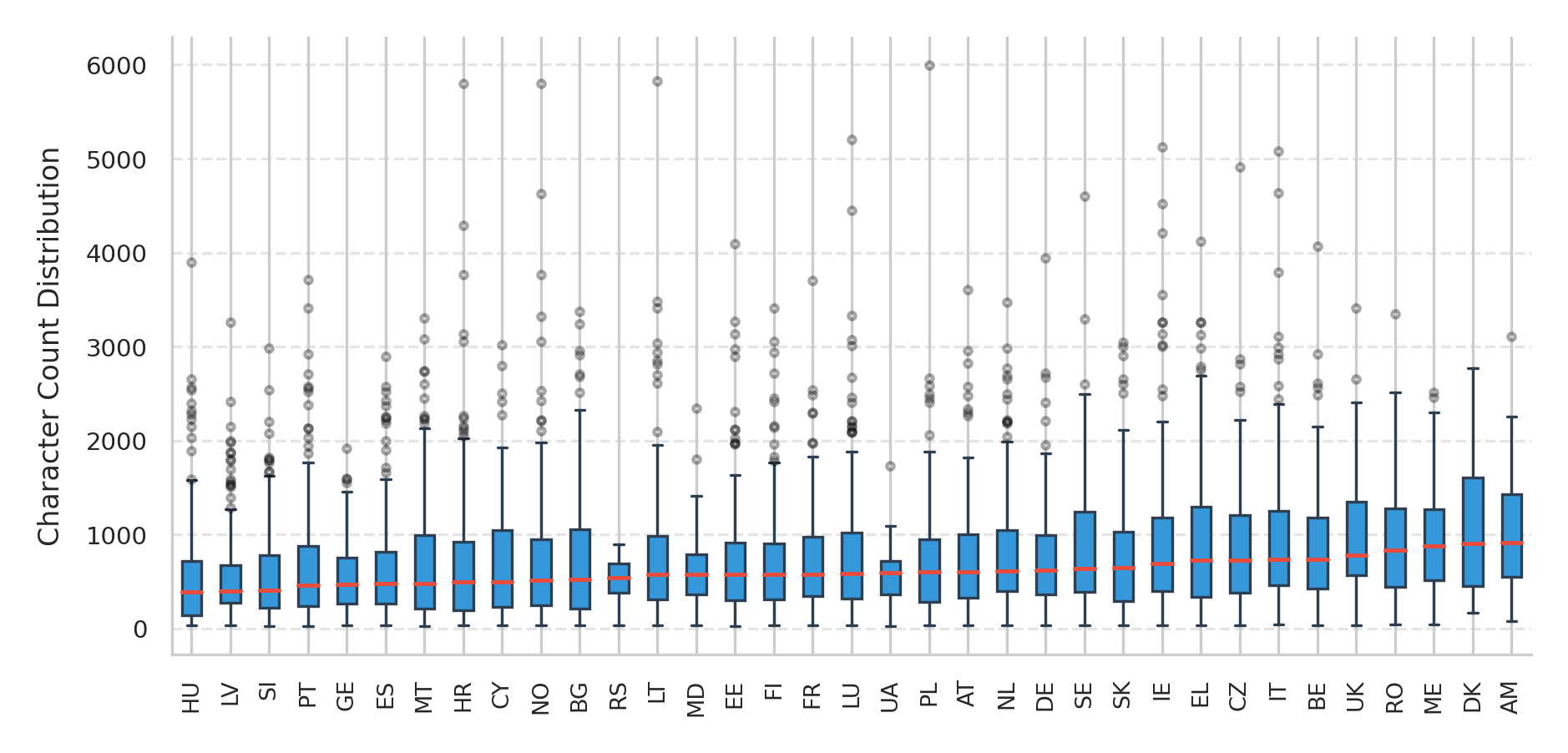}
    \caption{Boxplot of content length by country (ISO codes), sorted by median.}
    \label{fig:country_len}
  \hfill
\end{figure}
Figure \ref{fig:country_len} presents the boxplot distribution of character counts per country across the reporting period (2012–2024), sorted by median length. We observe cross-country heterogeneity: while some countries maintained relatively concise reports, such as Hungary (HU) and Latvia (LV), others consistently produced longer textual descriptions, such as Armenia (AM) and Denmark (DK). Several countries, such as Poland (PL) or Norway (NO), show individual outlier entries around 6,000 characters, indicating that while most reports for these countries are concise, a small number of sections are reported in considerably greater depth.

\begin{figure}[htbp]
    \centering
    \includegraphics[width=0.8\linewidth]{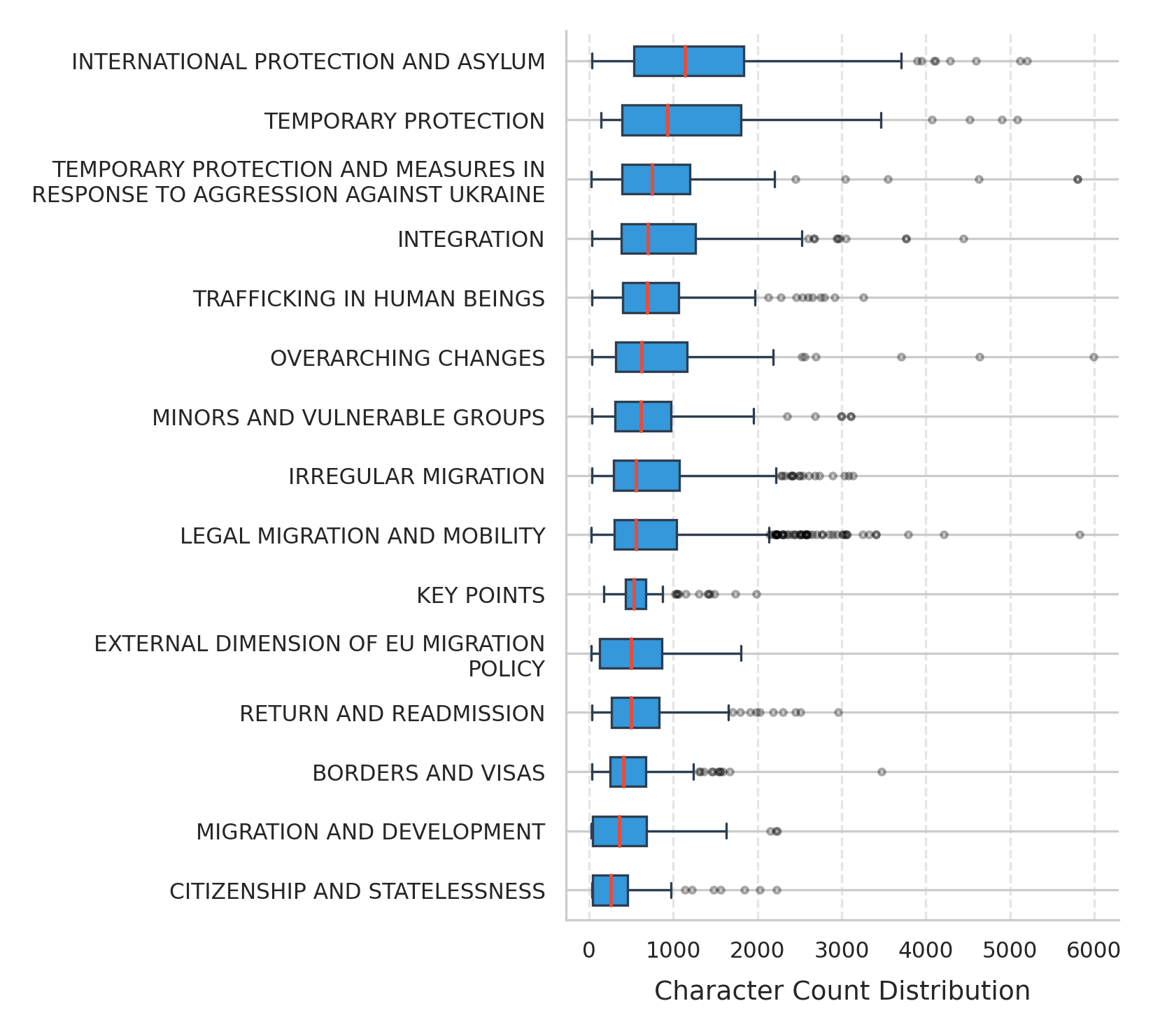}
    \caption{Content length distribution by section.}
    \label{fig:sectiond_len}
  \hfill
\end{figure}
Figure \ref{fig:sectiond_len} presents the character count distribution per section, sorted by median. A clear hierarchy emerges: protection-related sections exhibit the highest median character counts and the widest interquartile ranges, reflecting both greater narrative elaboration and greater variation in how countries report on these topics. In contrast, domains such as Citizenship and Statelessness and Migration and Development show the shortest medians. Across nearly all sections, the distributions are right-skewed, with a long tail of outlier entries. This is consistent with the fact that migration reporting occasionally requires substantially more detail for specific developments (e.g., a new asylum policy or a legislative reform) than the typical entry.
These structural differences demonstrate that the dataset captures policy domains with varying degrees of administrative complexity and narrative elaboration, reinforcing its suitability for comparative and longitudinal text analysis.


\FloatBarrier
\subsection{Lexical Trend Analysis}
We examine one domain subject to intense policy scrutiny and public debate: irregular migration. This basic analysis shows how the dataset can be used to track discursive shifts and policy priorities across Europe over a 13-year period (2012–2024).

We first extracted the most frequent unigrams and bigrams using a CountVectorizer, removing stopwords, country names, years, and generic administrative terms. We then mapped these high-frequency terms against the official EMN subsection taxonomy. 

As shown in Table \ref{tab:cluster}, we defined four thematic clusters. To ensure a balanced lexical comparison, each cluster is represented by five keywords drawn directly from the most frequent terms in the corpus. These clusters are intended as interpretable dimensions grounded in the institutional structure of the EMN factsheets. Frequencies were normalized per 1,000 words to account for variations in report length across different Member States and years.


\begin{table}[!htp]
\centering
\small
\begin{tabular}{p{1.3cm} p{2.5cm} p{2.8cm}}
\toprule
\textbf{EMN-section} & \textbf{Cluster} & \textbf{Keywords} \\
\midrule

& Return Procedures 
& return, returns, voluntary return, 
  readmission, reintegration \\[0.2cm]

& Border Management \& Prevention 
& border, border control, control, 
  prevent, security \\[0.2cm]
Irregular Migration 
& Smuggling \& Criminal Enforcement 
& smuggling, trafficking, criminal, 
  police, officers \\[0.2cm]

& Institutional Cooperation \& Governance 
& cooperation, exchange, joint, 
  agreement, information campaign \\
\hline

\end{tabular}

\caption{Keyword clusters used in the lexical analysis of the dataset.}
\label{tab:cluster}
\end{table}

\begin{figure}
  \centering

    \includegraphics[width=0.8\linewidth]{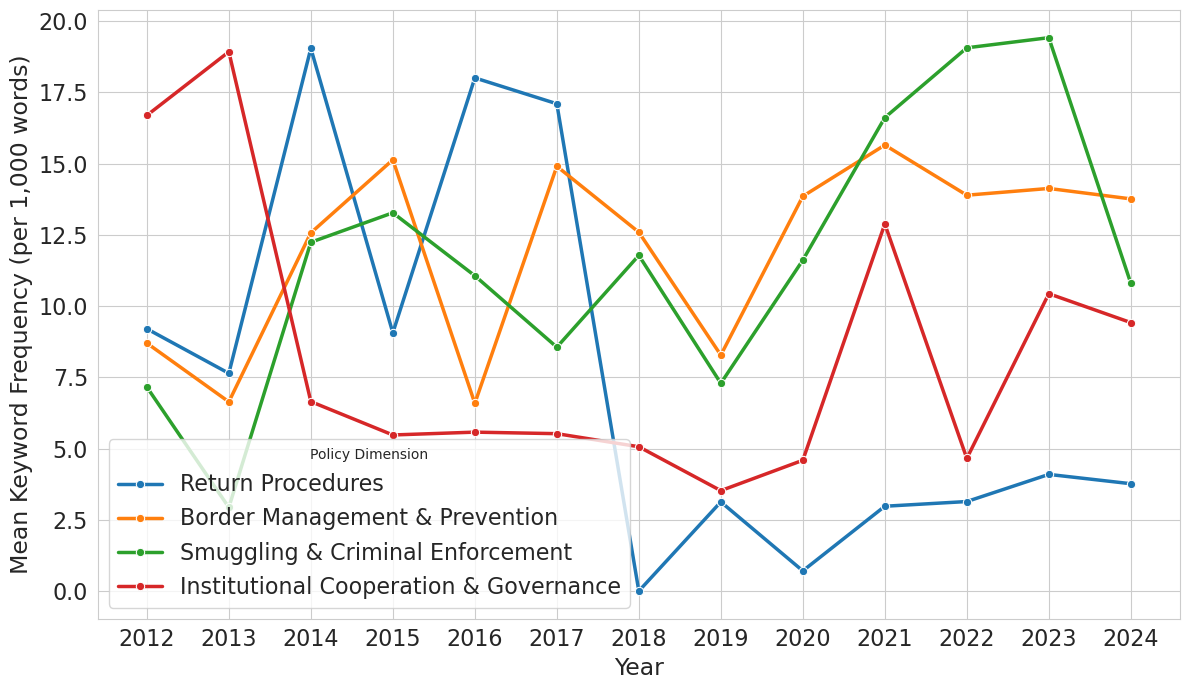}

  \caption{Evolution of Thematic Terminology in the Irregular Migration section for the 2012-2024 period and 35 European countries.}
  \label{fig:clusters}
\end{figure}

Figure \ref{fig:clusters} presents the evolution of thematic terminology within the irregular migration section (2012–2024). The plot reveals distinct phases of policy emphasis. \textit{"Return Procedures" }(e.g., voluntary return, readmission) exhibit a marked peak between 2014 and 2017. This spike coincides with the institutional aftermath of the 2015 migratory crisis, during which return became a primary focus of EU-level policy solutions \cite{european2015refugee}. In contrast, the terminology of \textit{"Border Management \& Prevention"} remains comparatively stable throughout the period, suggesting a sustained, baseline emphasis on control and prevention. On the other hand, the terminology of \textit{"Smuggling and Criminal Enforcement"}  shows a gradual increase, reaching its highest levels in the 2022–2023 period. This likely reflects legislative shifts and the emergence of new \textit{"anti-smuggling"} action plans at the EU level \cite{europejska2021renewed}. Finally, the \textit{"Institutional Cooperation and Governance"} cluster shows a distinct "U-shaped" trajectory. It is most prominent in the early years of the dataset (2012–2013), suggesting an initial period of intense administrative framing and the establishment of bilateral frameworks. Following a period of relative decline, the cluster shows a notable resurgence in 2021. This trend reflects the renewed focus on multilateral management and information-sharing protocols in the wake of shifting migratory routes.

While this analysis does not establish causal explanations, it illustrates the dataset’s capacity to capture longitudinal shifts in discursive emphasis. We want to highlight the analytical potential of structured, section-based corpus analysis within the EMN reports, arguing that the same could be done with other policy domains such as migrant integration or legal migration.

\section{Potential applications and uses}
\label{sec:potential}
The curated dataset allows the content of the EMN factsheets from 2012 to 2024 to be contained in a single file, enabling different uses. Below, we provide a few examples.
\begin{itemize}
    \item \textbf{Interactive dashboards.} The dataset could be visualized through dashboards where migration experts can compare the different policies applied across countries and years, instead of having to open several PDF files to make that kind of comparison or analysis. 
    \item \textbf{Chatbots.} The dataset could be integrated into RAG systems where citizens or experts could ask about different policies in Europe regarding migration, limiting hallucinations by constraining the answers to the content of the dataset. This would enable interested individuals (citizens or researchers) to be better informed, facilitating access to resources considered reliable sources that can be consulted when questions arise while reading social media or newspapers or conducting research on data science or media studies, thus serving as an additional source of credible information.
    
    \item \textbf{Fact checking.} As the data comes from European governments and it is considered a reliable resource, it could be used to extract claims to conduct fact checking by journalists, as a way to combat misinformation that may exist around this topic.
    \item \textbf{NLP analysis.} As examples, the dataset could allow Cross-National Lexical and Terminology Variation studies (in Computational Linguistics) to automatically identify country-specific or year-specific jargon for common migration concepts (e.g., different national terms used for \textit{"trafficking victim"} or \textit{"irregular migrant"}.

\end{itemize}

\section{Ethics and FAIR principles}
\label{sec:fair}
The dataset contains no personally identifiable information (PII), as it is derived exclusively from public institutional reports. The resource adheres to FAIR principles: it is \textbf{\textit{Findable}} via a permanent DOI (\url{https://doi.org/10.34777/fxzr-nc56})
, \textbf{\textit{Accessible}} through a public repository
\url{https://www.idiap.ch/dataset/emn-factsheets}, \textbf{\textit{Interoperable}} via standard CSV format
, and \textbf{\textit{Reusable}} under a CC BY 4.0 license.

\section{Conclusions}
\label{sec:conclusions}
In this paper, we presented a publicly available dataset containing the structured textual content of EMN factsheets, which provide annual overviews on the main developments and policies implemented in the area of migration across 35 countries over a 13-year period (2012-2024). This data is designed to support evidence-based policy analysis and computational social science, allowing researchers to benchmark national policy trajectories and track the evolution of institutional narratives. We illustrated the value of the dataset through descriptive analyses of the corpus, including variations in the length of contributions across countries and thematic sections. Additionally, we showed concrete examples of temporal patterns in a specific migration policy domain. Finally, by providing a structured and reliable primary source, this dataset has several possible uses related to the verification of policy-related claims, offering a resource to potentially counter online misinformation regarding migration with documented institutional facts.
\begin{acks}
This work was supported by the ELIAS project, funded by the European Commission (Grant 101120237). We would like to thank the European Migration Network (EMN) for providing us with factsheets from previous years that are no longer publicly available.
\end{acks}
\section*{GenAI Usage Disclosure}
Generative AI tools were used to assist with coding for this project.

\bibliographystyle{ACM-Reference-Format}
\bibliography{sample-base}


\end{document}